\documentclass[camera,letterpaper,nomarginnotes,nonarrowgutter, 11pt]{jpaper}

\usepackage[table]{xcolor}
\usepackage[T1]{fontenc}
\usepackage{booktabs}
\usepackage{listings}
\usepackage{fancyhdr}
\usepackage{datetime}
\usepackage{cite}

\usepackage{textcomp}
\usepackage{breakurl}

\usepackage{bm}
\makeatletter
\AtBeginDocument{\DeclareMathVersion{bold}
\SetSymbolFont{operators}{bold}{T1}{times}{b}{n}
\SetMathAlphabet{\mathrm}{bold}{T1}{times}{b}{n}
\SetMathAlphabet{\mathit}{bold}{T1}{times}{b}{it}
\SetMathAlphabet{\mathbf}{bold}{T1}{times}{b}{n}
\SetMathAlphabet{\mathtt}{bold}{OT1}{pcr}{b}{n}
\SetSymbolFont{symbols}{bold}{OMS}{cmsy}{b}{n}
\renewcommand\boldmath{\@nomath\boldmath\mathversion{bold}}}
\makeatother

\def\BibTeX{{\rm B\kern-.05em{\sc i\kern-.025em b}\kern-.08em
    T\kern-.1667em\lower.7ex\hbox{E}\kern-.125emX}}

\usepackage{url}

\usepackage{setspace}
\usepackage[hang]{footmisc}

\usepackage{pdfpages}

\usepackage{graphbox,graphicx}\usepackage{subcaption}
\usepackage{tabularx}
\usepackage{adjustbox}
\usepackage{rotating}
\usepackage{multirow}
\usepackage{booktabs}

\usepackage{algorithm}
\usepackage[noend]{algpseudocode}
\usepackage{listings}

\usepackage[colorlinks = true,
            linkcolor = blue,
            urlcolor  = blue,
            citecolor = blue,
            anchorcolor = blue,
            breaklinks
            ]{hyperref}

\usepackage[resetlabels]{multibib}

\usepackage{xspace}

\usepackage{fancyhdr}
\usepackage{balance}

\newcommand{\sequencingTechnologyReviewPapers}[0]{
    mccombie2019next,
    wang2021nanopore,
    logsdon2020long,
    giani2020long%
}

\newcommand{\dtwSeminal}[0]{
    velichko1970dtw,
    sakoe1970dtw,
    sakoe1971dtw%
}

\newcommand{\citesignalanalysis}{bao_squigglenet_2021,loose_real-time_2016,zhang_real-time_2021,kovaka_targeted_2021,senanayake2023deepselectnet,sam_kovaka_uncalled4_2024,firtina2023rawhash,firtina_rawhash2_2024,shih_efficient_2022,sadasivan_rapid_2023,dunn2021squigglefilter,shivakumar_sigmoni_2024,sadasivan_accelerated_2024,gamaarachchi2020dtwgpu,samarasinghe2021dtwfpga}

\definecolor{darkblue}{rgb}{0,0,0.8}
\definecolor{lightorange}{rgb}{1.0,0.8,0.6}
\definecolor{beige}{rgb}{0.835,0.741,0.686}
\definecolor{rose}{rgb}{1.0,0.8,0.9}
\definecolor{darkpink}{rgb}{1.0,0.2,0.4}

\newcommand{\fulltitle}[0]{RawAlign: Accurate, Fast, and Scalable Raw Nanopore Signal Mapping via Combining Seeding and Alignment}

\newcommand{\toolname}[0]{RawAlign\xspace}

\newcommand{\gitrepo}[0]{\url{https://github.com/CMU-SAFARI/RawAlign}}

\newcommand{\x}[0]{$\times$\xspace}

\renewcommand{\algref}[1]{{\hyperref[alg:#1]{Alg.~\ref*{alg:#1}}}}
\newcommand{\alglineref}[1]{{\hyperref[algline:#1]{Line~\ref*{algline:#1}}}}
\newcommand{\algrangeref}[2]{{\hyperref[algline:#1]{Lines~\ref*{algline:#1}\nobreakdash-\ref*{algline:#2}}}}
\newcommand{\secref}[1]{\hyperref[sec:#1]{\S\ref*{sec:#1}}}
\newcommand{\figref}[2][]{{\hyperref[fig:#2]{Fig.~\ref*{fig:#2}#1}}}
\newcommand{\tabref}[1]{{\hyperref[table:#1]{Tab.~\ref*{table:#1}}}}
\newcommand{\theoref}[1]{{\hyperref[theorem:#1]{Theorem~\ref*{theorem:#1}}}}

\newcommand{\supsecref}[1]{\hyperref[supsec:#1]{Sup.\S\ref*{supsec:#1}}}
\newcommand{\supfigref}[2][]{{\hyperref[supfig:#2]{Sup.Fig.~\ref*{supfig:#2}#1}}}
\newcommand{\suptabref}[1]{{\hyperref[suptable:#1]{Sup.Tab.~\ref*{suptable:#1}}}}
\newcommand{\suptheoref}[1]{{\hyperref[suptheorem:#1]{Sup.Theorem~\ref*{suptheorem:#1}}}}

\newcommand{\ru}[0]{Read Until\xspace}

\newcommand{\uncalled}[0]{UNCALLED\xspace}
\newcommand{\rawhash}[0]{RawHash\xspace}
\newcommand{\rawhashtwo}[0]{RawHash2\xspace}
\newcommand{\sigmap}[0]{Sigmap\xspace}
\newcommand{\minimap}[0]{minimap2\xspace}

\newcommand{\defn}[1]{\textit{#1}}

\newcommand\rev[1]{{\color{black}{#1}}}

\title{\fulltitle}
\newcommand{\affilETH}[0]{\textsuperscript{\S}}

\newcommand{\authorgap}[0]{\hspace{3em}}
\author{
{Joël Lindegger\affilETH}\authorgap%
{Can Firtina\affilETH}\authorgap%
{Nika Mansouri Ghiasi\affilETH}\\%
{Mohammad Sadrosadati\affilETH}\authorgap%
{Mohammed Alser\affilETH}\authorgap%
{Onur Mutlu\affilETH}
\vspace{2mm}\\%
\emph{\affilETH ETH Zurich}
}

\newcites{supp}{Supplementary References}

\begin{document}
\bstctlcite{IEEEexample:BSTcontrol}

\maketitle
\thispagestyle{plain}

\begin{abstract}Nanopore sequencers generate raw electrical signals representing the contents of a biological sequence molecule passing through the nanopore. These signals can be analyzed directly, avoiding basecalling entirely.
We observe that while existing proposals for raw signal analysis typically do well in all metrics for small genomes (\emph{e.g.,} viral genomes), they all perform poorly for large genomes (\emph{e.g.,} the human genome).
\textbf{Our goal} is to analyze raw nanopore signals
in an accurate, fast, and scalable manner.
To this end, we propose \toolname, the first work to integrate fine-grained signal alignment into the state-of-the-art raw signal mapper. To enable accurate, fast, and scalable mapping with alignment, \toolname implements three algorithmic improvements and hardware acceleration via a vectorized implementation of fine-grained alignment. Together, these significantly reduce the overhead of typically computationally expensive fine-grained alignment.
Our extensive evaluations on different use cases and various datasets show \toolname provides 1)~the most accurate mapping for large genomes and 2)~and on-par performance compared to \rawhash (between 0.80\x-1.08\x), while achieving better performance than \uncalled and \sigmap by on average (geo. mean) 2.83\x and 2.06\x, respectively.
\noindent\textbf{Availability:} \gitrepo
\end{abstract}

\section{Introduction} \label{sec:introduction}
Nanopore-based sequencers generate a series of raw electrical signal measurements representing the contents of a biological sequence molecule passing through the sequencer's nanopore.
Nanopore sequencers provide unique features and, thus, pose unique challenges to computational analysis~\cite{\sequencingTechnologyReviewPapers}.
For example, sequencers can provide real-time control over the sequencing of individual molecules with a mechanism called \defn{\ru}~\cite{ontminion}. \ru ejects a molecule from its pore before it is fully sequenced if deemed irrelevant (\emph{e.g.,} by computational methods)~\cite{loose_real-time_2016}. 
The challenge to efficiently leveraging such real-time control is to (1)~make \emph{timely} decisions (otherwise, sequencing time is wasted), (2)~make \emph{accurate} decisions (otherwise, relevant data may be lost, or data may be biased), (3)~match at least the \emph{throughput} of the sequencer (otherwise, the computation is not real-time)~\cite{dunn2021squigglefilter,firtina2023rawhash,firtina_rawhash2_2024}.

Although a common approach for real-time analysis starts with translating raw nanopore signals into nucleotide characters with the \emph{basecalling} step, a number of recent works enable analyzing raw nanopore signals directly without basecalling\rev{~\cite{mutlu2023genomicscodesign,sam_kovaka_uncalled4_2024}} to enable a faster and power-efficient analysis (\emph{e.g.,}~\cite{zhang_real-time_2021,kovaka_targeted_2021,firtina2023rawhash,firtina_rawhash2_2024,dunn2021squigglefilter,shih_efficient_2022,sadasivan_accelerated_2024,gamaarachchi2020dtwgpu,samarasinghe2021dtwfpga,bao_squigglenet_2021,noordijk2023baseless,senanayake2023deepselectnet,sadasivan_rapid_2023}). Such an analysis can mainly be useful for better scalability and portable sequencing in
resource-constrained devices~\cite{dunn2021squigglefilter}. These earlier works already enable important applications such as relative abundance calculation, contamination testing, and real-time sequencing decisions while completely forgoing the need for basecalling (\emph{i.e.,} converting the raw signals to the nucleotide bases A, C, G, and T).
Among these works, the state-of-the-art work~\cite{firtina2023rawhash, firtina_rawhash2_2024} enables mapping these raw nanopore signals to a reference genome using a quick hash-based search followed by chaining. Such a coarse-grained sequence comparison is fast and scales well to large reference genomes, but (1)~has relatively low accuracy, and (2)~lacks fine-grained sequence alignments in the output, limiting its applicability to downstream analysis tasks (\emph{e.g.,} variant calling).

\textbf{Our goal} is to analyze raw nanopore signals in an accurate, fast, and scalable manner.
To this end, we propose \defn{\toolname}, the first tool to combine fast and scalable hash-based raw signal mapping algorithms with highly accurate alignment \rev{based on dynamic time warping (DTW)~\cite{\dtwSeminal}} to enable better accuracy and performance, while allowing future work to focus on further downstream analysis after mapping using the fine-grained alignment information (\emph{e.g.,} variant calling).

However, combining DTW with a hash-based raw signal mapper to enable scalable and accurate analysis is not trivial as we identify three key \textbf{challenges}. First, DTW is a computationally costly algorithm that prevents the earlier works that use DTW from scaling to perform real-time analysis for larger genomes~\cite{dunn2021squigglefilter,sadasivan_accelerated_2024,shih_efficient_2022}. Second, hash-based raw signal mappers tend to generate many candidate regions based on hash matches, resulting in frequent calls to the DTW subroutine, exacerbating DTW's already high computational cost. Third, DTW's scoring mechanism is not directly suitable for integration with the mapping decision process because it depends on variables such as path length and cumulative distance, which may not accurately reflect the optimal alignment or the best mapping quality in the context of raw signal data.

To tackle these challenges, \toolname efficiently integrates DTW into a hash-based raw signal mapper by designing and integrating optimizations when calculating DTW and improving the decision-making process in four directions:
\begin{itemize}
    \item The number of candidate locations is reduced with \rawhash's chaining-based filtering (see \secref{seed-filter-align})
    \item The number of calls to the DTW subroutine is further reduced via an early termination strategy (see \secref{early_termination})
    \item The number of arithmetic operations for the DTW computation is reduced via \defn{anchor-guided alignment} (see \secref{sparse_alignment}) and \defn{banding}/\defn{windowing}  (see \secref{banded_alignment})
    \item The arithmetic operations are implemented efficiently through \rev{Single Instruction, Multiple Data (SIMD)}~\cite{flynn_very_1966} vectorization (see \secref{simd})
\end{itemize}
In combination, this strategy makes DTW efficient and scalable to large genomes, enabling \toolname to analyze raw nanopore signals efficiently and accurately for a wide range of reference genome sizes.

We make the following contributions:
\begin{itemize}
    \item \rev{We develop \toolname, the first tool for raw signal mapping to combine seeding and alignment.}
    \item \rev{We comprehensively evaluate \toolname and demonstrate its applicability to a wide range of datasets. \toolname provides similar throughput to \rawhash while consistently improving accuracy.}
    \item We develop optimization techniques that collectively make practical the combination of seeding and alignment for raw nanopore signal analysis.
    \item We demonstrate that \toolname is the first tool to map raw nanopore signals to large reference genomes with high accuracy.
    \item We demonstrate the generality of \toolname by integrating it with and comprehensively comparing it to RawHash2.
    \item We open-source all code of \toolname and associated evaluation methodology, available at \gitrepo. \toolname can be readily used as a raw nanopore signal mapper.
\end{itemize}

\section{Methods} \label{sec:methods}
\begin{figure}[htb]
  \centering
    \includegraphics[width=\linewidth]{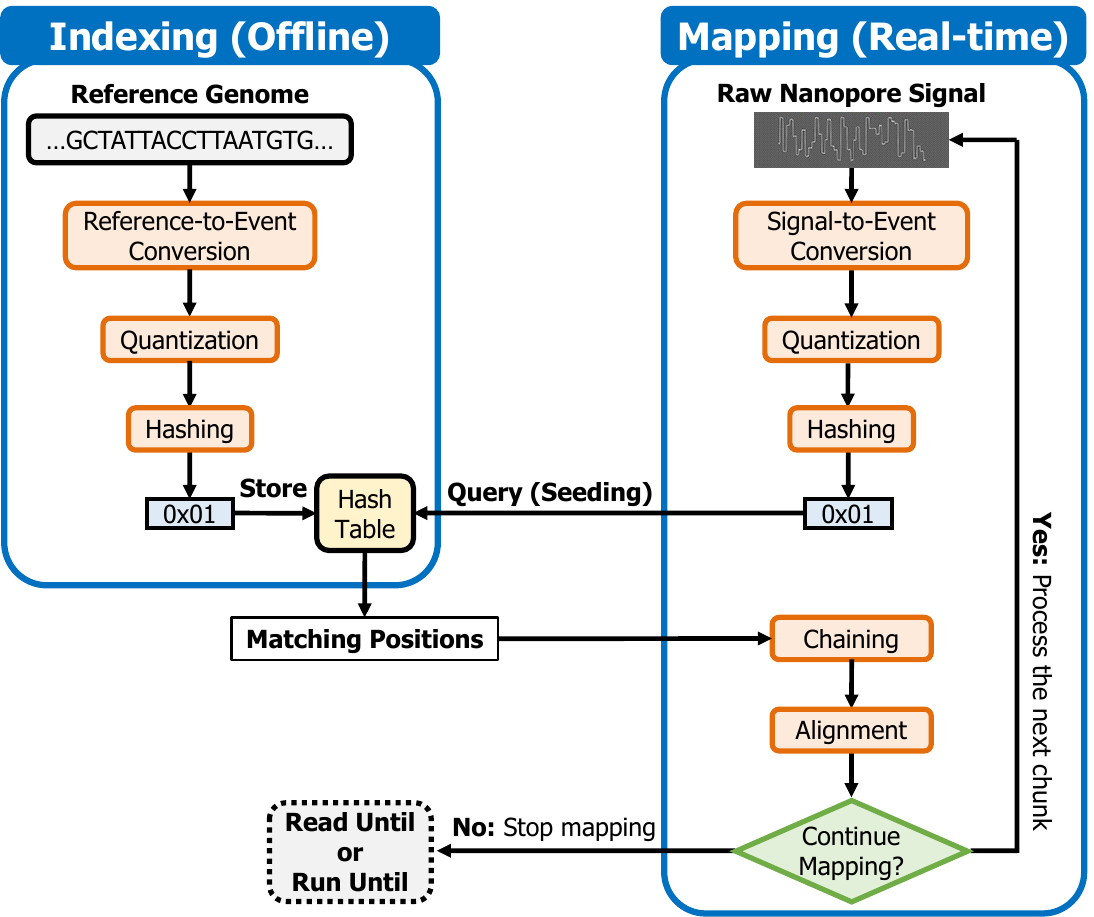}
  \caption{Overview of \toolname.}
  \label{fig:overview}
\end{figure}

To identify the mapping positions and the alignment between a reference genome and a raw nanopore signal, \toolname works in three major steps: (1)~offline reference database pre-processing (see~\supsecref{ref_preprocessing}), (2)~online read pre-processing (see~\supsecref{read_preprocessing}), and (3)~online mapping of the reads with the Seed-Filter-Align paradigm~(see~\secref{seed-filter-align}). \rev{\figref{overview} shows the overview of the steps that \toolname takes to perform the mapping.} \toolname uses the seeding and filtering methods from the prior work \rawhash~\cite{firtina2023rawhash}. We explain the alignment based on dynamic time warping (DTW) in detail in~\secref{dtw}, and we explain how \toolname uses the result from DTW for making mapping decisions in~\secref{mappingdecisions}. To make the Seed-Filter-Align paradigm practical for raw signal mapping, we introduce four strategies to improve the performance of the alignment step in \secref{perfopt}.

\subsection{Mapping via Seed-Filter-Align} \label{sec:seed-filter-align}
\toolname follows the \defn{Seed-Filter-Align} or \defn{Seed-Filter-Extend} paradigm, which has been applied with great success by conventional basecalled read mappers (\emph{e.g.,}~\cite{mini2-paper,li2013bwamem,vasimuddin2019bwamem2}). The main mapping process consists of three steps: \defn{seeding}, \defn{filtering}, and \defn{alignment}. \figref{alignmentexample} shows a real example of each of these steps in \toolname for a read in the d2 \emph{E. Coli} dataset. 

\subsubsection{Seeding}\label{sec:seeding}
\begin{figure}[H]
  \centering
    \includegraphics[width=\linewidth]{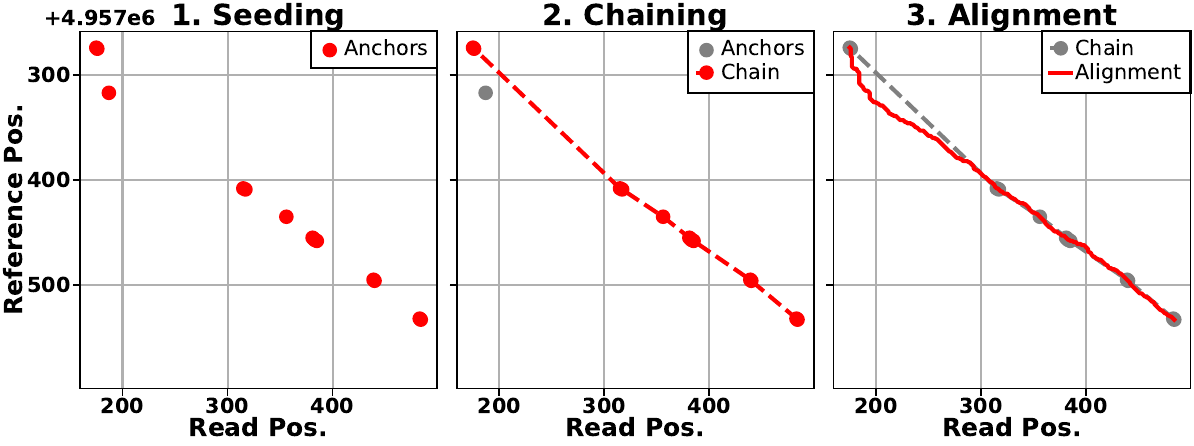}
  \caption{Example of \toolname's Usage of Seed-Filter-Align.}
  \label{fig:alignmentexample}
\end{figure}

First, during seeding, \toolname finds a large number of possible candidate locations
via fuzzy matches between the read and the reference database (\figref{alignmentexample}, left). Seeding is relatively fast and scales well to large reference genomes. \rev{To find these seed matches, \toolname uses the seeding step of \rawhash~\cite{firtina2023rawhash}, which works in three steps as shown in \figref{overview}. First, it converts both the reference genome and raw nanopore signals into a discrete set of signals representing k-mers, called \defn{events}. Second, to reduce the residual noise in these events (\emph{i.e.,} due to certain imperfections in sequencing and raw signal-to-discrete signal conversion), RawHash quantizes these events and generates hash values from these quantized values, called \defn{seeds}.} Third, the locations where each seed occurs in the reference database are obtained by querying the index with the seed. The read-reference location pairs are called \defn{anchors}.

\subsubsection{Filtering}\label{sec:filtering}
Second, during filtering, \toolname reduces the number of anchors by heuristically evaluating and removing those that do not look promising. Filtering is a relatively coarse-granular comparison between sequences, and hence, faster than DTW-based alignment but less accurate on its own. One class of filters is \defn{chaining} \rev{(as we show in the middle plot in \figref{alignmentexample} and in \figref{overview}).} During chaining, the relative positions of anchors are evaluated with a dynamic programming algorithm to identify a set of anchors that is approximately colinear in the read and reference sequence. \toolname uses the same chaining logic as \rawhash~\cite{firtina2023rawhash} and \sigmap~\cite{zhang_real-time_2021}.

\subsubsection{Alignment}\label{sec:alignment}
Third, during \defn{alignment} or \defn{extension}, \toolname re-evaluates all remaining candidate regions after chaining \rev{(see \figref{overview})} at fine granularity using dynamic time warping (DTW) (\figref{alignmentexample}, right). This finely granular comparison significantly improves accuracy relative to prior works that avoid it (\emph{e.g.,}~\cite{firtina2023rawhash,firtina_rawhash2_2024}). We elaborate in \secref{dtw} why DTW~\cite{\dtwSeminal,dunn2021squigglefilter,sadasivan_accelerated_2024,shih_efficient_2022} is a natural choice to evaluate candidate locations, what its challenges are, and how \toolname overcomes them.

\subsection{Alignment via Dynamic Time Warping} \label{sec:dtw}
\rev{To perform a pairwise comparison and alignment of \emph{basecalled} sequences, \defn{Needleman-Wunsch (NW)} is a widely used algorithm~\cite{needleman1970general}. However, these commonly used algorithms for basecalled sequences are \emph{not} suitable for aligning signals as the length of the signals can vary due to the varying speed at which biological molecules move through the nanopore during sequencing, known as \defn{translocation speed}. Variations in translocation speed result in time distortions in the signal data. This makes raw signal analysis a time-series analysis to handle this time variance.}

\begin{figure}[H]
  \centering
    \includegraphics[width=\linewidth]{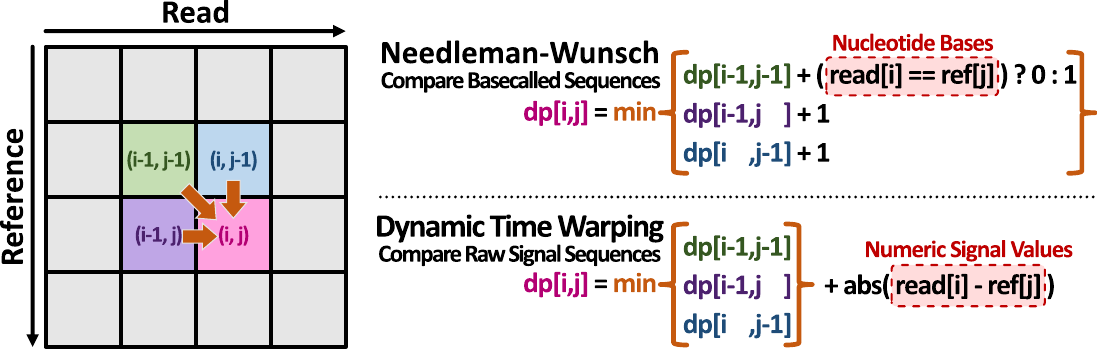}
  \caption{Conceptual Similarity of Needleman-Wunsch and Dynamic Time Warping.}
  \label{fig:nw_dtw_comparison}
\end{figure}

\defn{Dynamic time warping (DTW)} is an algorithm for fine-grained pairwise comparison and alignment of time series data~\cite{\dtwSeminal}, such as two raw nanopore signals. Multiple prior works use DTW to compare raw nanopore signals to entire reference genomes (\emph{e.g.,} SquiggleFilter~\cite{dunn2021squigglefilter}, DTWax~\cite{sadasivan_accelerated_2024}, and HARU~\cite{shih_efficient_2022}). \rev{To show how DTW can handle the time variance in signals compared to NW while identifying similarities between a pair of signals (as in NW), we show the differences and similarities between NW and DTW in \figref{nw_dtw_comparison}.}

\rev{The key similarity between NW and DTW is that both are based on dynamic programming,} filling a two-dimensional table of numbers. Each entry in the dynamic programming table is computed based on three neighbor entries in the north, northwest, and west according to simple update rules. The final alignment score (for NW) or cost (for DTW) is the entry in the southwest corner of the table.

There are two key differences between NW and DTW: First, NW's update rule performs exact character comparisons, suitable for aligning basecalled sequences where nucleotides either match or do not match (\emph{e.g.,} A$=$A, A$\neq$G). In contrast, DTW computes the numerical difference between signal values, which accommodates the continuous and noisy nature of raw signal data (\emph{e.g.,} 0.2$\approx$0.3, but 0.1$\neq$0.9). Second, NW imposes fixed penalties for insertions and deletions (\defn{indels}), independent of the sequence content, which is appropriate for sequences with rare indels. DTW, however, adjusts penalties based on the similarity of the signal values, effectively handling frequent indels caused by variable translocation rates and segmentation imperfections in raw nanopore signals (see~\supsecref{read_preprocessing}).

\subsection{Alignment Score and Mapping Decisions} \label{sec:mappingdecisions}
\toolname bases its mapping decisions \rev{(Continue Mapping part in \figref{overview})} mainly on the alignment (DTW) cost (see \secref{dtw}). The DTW cost is always positive, where a low cost indicates a good match between the read and the reference, thus low-cost candidate regions should generally be preferred. However, the DTW cost monotonically increases with sequence length, meaning that a long, well-matching candidate region typically has a higher DTW cost than a short, well-matching candidate region. That is undesirable because a long, well-matching region intuitively should be preferred over a short one, as they indicate a better match between the read and the reference.

To address this issue, \toolname employs a customized alignment scoring scheme based on the DTW cost $C_{DTW}$ and a constant \defn{match bonus} $B_{Match}$ that is scaled by the number of aligned read events $N_{Read}$:
\vspace{-1em}
\[Score = B_{Match} \times N_{Read} - C_{DTW}\]
\vspace{-2em}

With this scoring scheme, longer well-matching candidate regions achieve higher scores than shorter ones, provided $B_{Match}$ is chosen appropriately. $B_{Match}$ should be chosen such that it is above the noise floor of the sequencer, \emph{i.e.,} such that two correctly aligned events usually differ by less than $B_{Match}$, or can be chosen using a parameter sweep.

Based on the alignment score, \toolname considers the highest scoring candidate location as mapped if its score is at least above some threshold $Score_{Threshold}$. $Score_{Threshold}$ is necessary to avoid false positives in the presence of the match bonus; thus, the two parameters need to be chosen in conjunction.

\subsection{Performance Optimizations to DTW} \label{sec:perfopt}
We observe, like prior work (\emph{e.g.,}~\cite{dunn2021squigglefilter,sadasivan_accelerated_2024,shih_efficient_2022}), that DTW can have a high computational overhead if not applied judiciously. Prior works partially solve this challenge by proposing hardware accelerators for DTW~\cite{dunn2021squigglefilter,sadasivan_accelerated_2024,shih_efficient_2022}. \toolname is, in principle, compatible with such accelerators and can benefit from them. However, to make it as out-of-the-box usable as possible, we optimize the DTW in \toolname at both the algorithm and software level. Since \toolname's mapping decisions require only the alignment score but not the exact alignment path, the exact path is not computed by default but can be added to the output file via a command line flag. We optimize our DTW implementation via a combination of four techniques: (1)~Early termination, (2)~anchor-guided alignment, (3)~banded alignment, and (4)~SIMD.

\subsubsection{Early Termination} \label{sec:early_termination}
Like other raw signal mappers, \toolname outputs at most one mapping location per read. We observe that this can be exploited to terminate DTW early using a branch-and-bound approach, where we prune computations that cannot yield better results than the current best solution. We can safely terminate the DTW computation for a candidate location as we can prematurely confirm if a candidate location's alignment score can surpass the current best alignment. This technique is most powerful if, by chance, the best candidate location with a very high alignment score is analyzed early, meaning all further alignments can be terminated more aggressively. To maximize the likelihood of the best candidate location being analyzed early, \toolname sorts candidate locations by their chaining score before alignment.

\subsubsection{Anchor-Guided Alignment} \label{sec:sparse_alignment}
\begin{figure}[htb]
  \centering
    \includegraphics[width=0.8\linewidth]{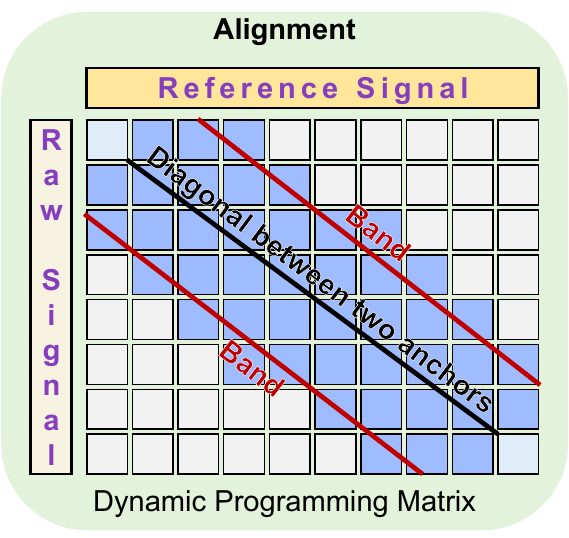}
  \caption{Anchor-Guided Alignment and Banded Alignment.}
  \label{fig:banding}
\end{figure}

We observe that the optimal alignment of a chain typically includes the anchors in the chain. Thus, it is sufficient to align only the gaps between anchors and then combine the individual alignments or scores. This improves the performance because we can avoid performing unnecessary alignment operations for signals that a single anchor covers \rev{(usually from tens of signals up to a few hundred signals). \figref{banding} shows such an alignment path (\emph{i.e.,} the diagonal path) between two anchors. Upper left and lower right cells in the dynamic programming matrix are anchors spanning a certain amount of signals. The alignment operations are performed without using the signals covered by these two cells in the matrix.} Similar techniques have been applied by conventional basecalled read mappers (\emph{e.g.,} \minimap~\cite{mini2-paper}) and can significantly reduce the number of dynamic programming cells that need to be computed. There can be a risk of inaccurate alignments if some of the anchors are not part of the globally optimal alignment; hence, \toolname provides command line options for either anchor-guided or global (\emph{i.e.,} non-anchor-guided) alignment. We observe that the accuracy loss due to anchor-guided alignment is typically small ($<1\%$ F-1 score on all evaluated datasets). Thus, \toolname uses anchor-guided alignment by default. 

\subsubsection{Banded Alignment} \label{sec:banded_alignment}
We observe that the optimal alignment of a chain or the gap between two anchors typically follows approximately the main diagonal (see \figref{alignmentexample} for an example and the diagonal path in \figref{banding}). This can be exploited by only computing a subset of the DTW dynamic programming table, called a \defn{band}. \rev{\figref{banding} shows the subset of an example of such a dynamic programming table where the alignment is performed only by using the cells that fall under or overlap with two bands. The distance of these bands to the main diagonal band is known as \defn{band width}.} Similar techniques have been applied by conventional basecalled read mappers and basecalled sequence aligners (\emph{e.g.,} Edlib~\cite{Edlib-btw753}, KSW2~\cite{ksw2_a}, \minimap~\cite{mini2-paper}). Banding can significantly reduce the number of arithmetic operations and memory accesses.

We observe that the read and reference sections of candidate pairs are rarely of the same length.
Therefore, our banding implementation adjusts the band to follow a diagonal with a slope of \( \frac{\text{read\_segment\_length}}{\text{reference\_segment\_length}} \) instead of the standard slope of 1, accommodating differences in sequence lengths between the read and reference segments.

\subsubsection{SIMD} \label{sec:simd}
DTW is a standard two-dimensional dynamic programming algorithm that can be accelerated with techniques such as SIMD, GPUs~\cite{sadasivan_accelerated_2024}, FPGAs~\cite{shih_efficient_2022}, or ASICs~\cite{dunn2021squigglefilter}.

We implement our banded dynamic time warping algorithm in an antidiagonal-wise~\cite{xia2022reviewswparallel} fashion, processing the dynamic programming matrix along its antidiagonals to enable efficient parallel computation. We further optimize the code to ensure that GCC v11.3.0 can auto-vectorize the critical innermost loop of the dynamic programming computation, which we verify via compilation logs.

\section{Results} \label{sec:results}
\begin{figure*}[htb]
  \centering
    \includegraphics[width=0.24\linewidth]{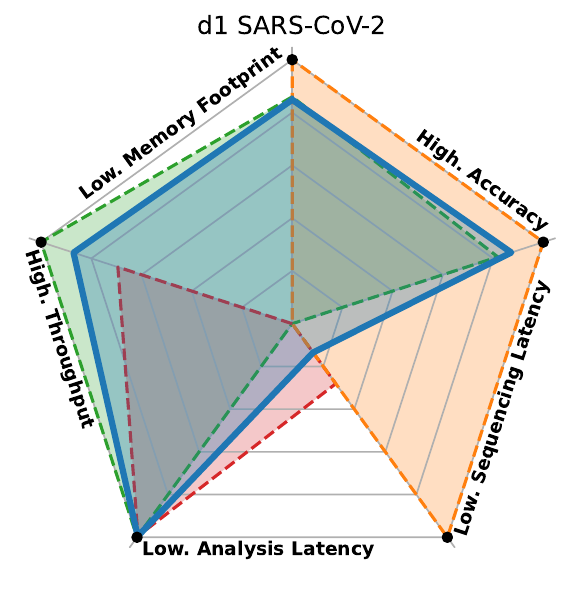}
    \hfill
    \includegraphics[width=0.24\linewidth]{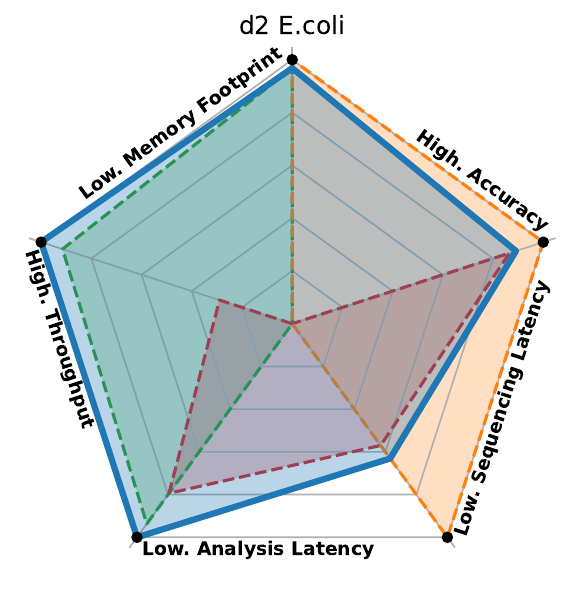}
    \hfill
    \includegraphics[width=0.24\linewidth]{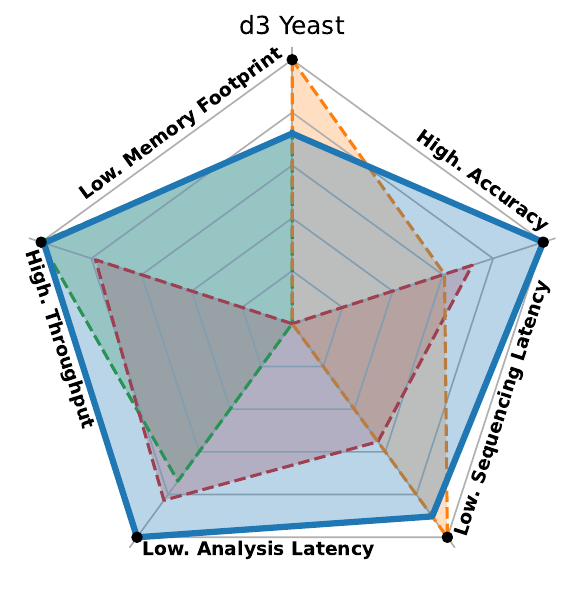}
    \hfill
    \includegraphics[width=0.24\linewidth]{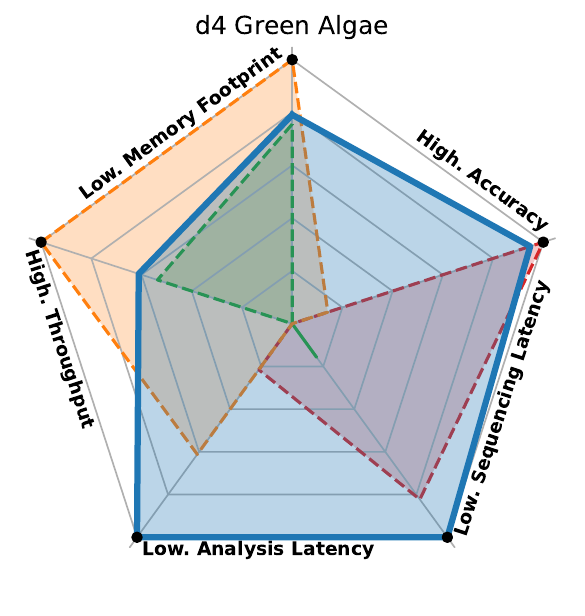}

    \includegraphics[align=c,width=0.15\linewidth]{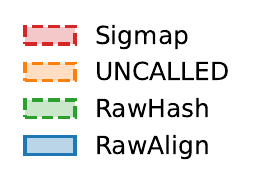}
    \includegraphics[align=c,width=0.24\linewidth]{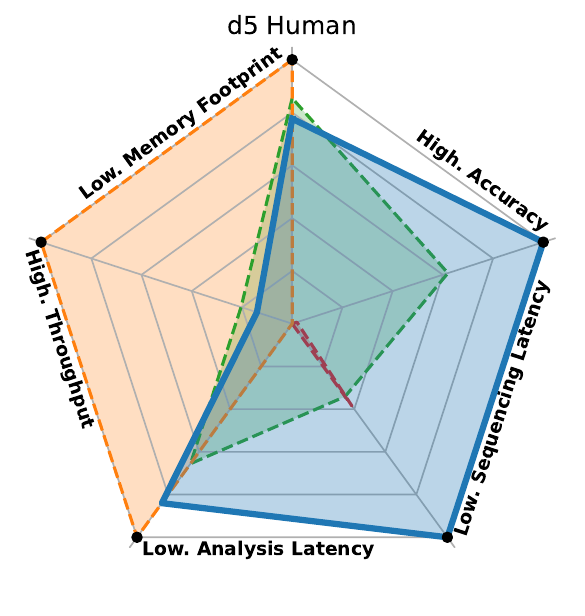}
    \hfill
    \includegraphics[align=c,width=0.24\linewidth]{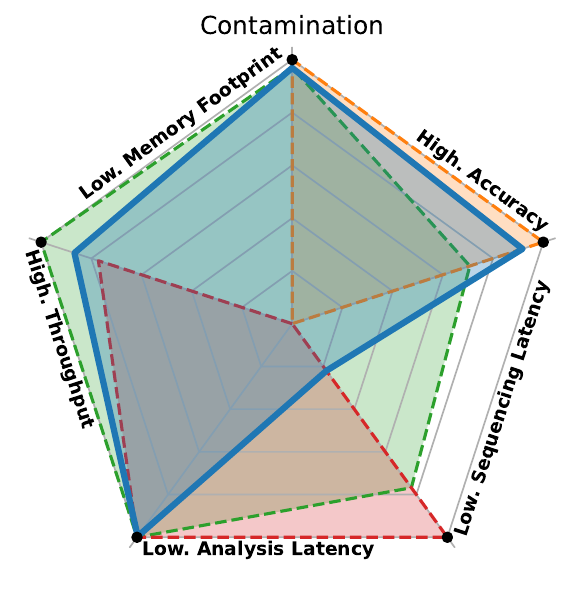}
    \hfill
    \includegraphics[align=c,width=0.24\linewidth]{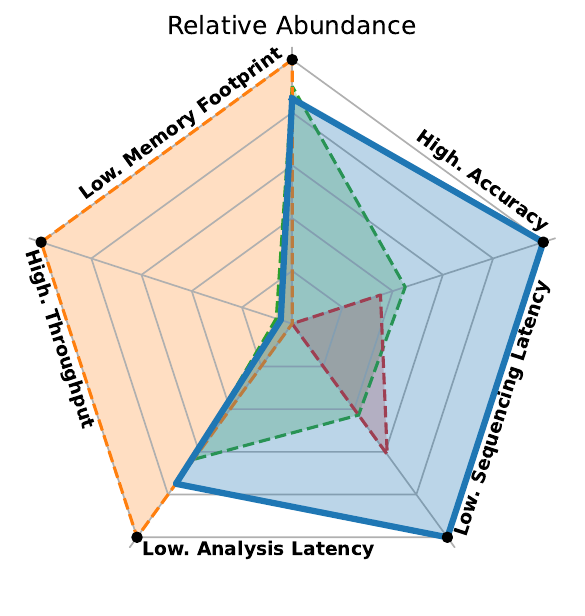}
  \caption{Accuracy-Latency-Throughput-Memory Footprint Tradeoffs of Rawalign \emph{vs.} Baselines.}
  \label{fig:spidertradeoffs}
\end{figure*}

\subsection{Evaluation Methodology}
\subsubsection{Baselines}
We demonstrate the benefits of \toolname by comparing to four prior works: \uncalled~\cite{kovaka_targeted_2021}, \sigmap~\cite{zhang_real-time_2021}, \rawhash~\cite{firtina2023rawhash}, and \rawhashtwo~\cite{firtina_rawhash2_2024}. We demonstrate the generality of \toolname by integrating its alignment components into \rawhashtwo without specific tuning. We refer to this configuration as \defn{\toolname-RH2}. \toolname and all four baselines are CPU-based. They accept a set of raw signal reads (\emph{e.g.,} FAST5 files) and a reference database (\emph{e.g.,} a FASTA file) to map to and produce mapping locations as a \rev{Pairwise mApping Format (PAF) file~\cite{li_minimap_2016}}. Due to a lack of a true ground truth in the read mapping experiments, we consider the mapping locations generated by \minimap~\cite{mini2-paper} as the ground truth and hence do not compare to its accuracy. In the relative abundance estimation evaluation, the true ground truth is known since the read set is artificially composed of reads from multiple species; thus, we compare the accuracy of the estimates to both \minimap and the raw signal baselines.

\subsubsection{Metrics}
We evaluate each tool using five categories of metrics:
\begin{itemize}
    \item \textbf{Memory footprint} (GB) during mapping as reported by Ubuntu's \texttt{time} package; lower is better. We also report the indexing memory footprint in \suptabref{full_results}.
    \item The mean \textbf{throughput} (bp/s) per mapping thread is crucial for real-time analysis. It should match or exceed the data generation rate of a nanopore sequencer, which is approximately 450 bp/s in recent \rev{Oxford Nanopore Technologies (ONT)} sequencers. Achieving this ensures that the computational analysis can keep up with the sequencing process; therefore, higher throughput is better. We also report the median throughput in \suptabref{full_results}.
    \item Mean \textbf{analysis latency} (ms), \emph{i.e.,} the time spent computing for mapping reads as reported by each tool in its PAF output; lower is better. We also report the median analysis latency in \suptabref{full_results}.
    \item Mean \textbf{sequencing latency} (chunks), where one \defn{chunk} corresponds to a segment of 450 base pairs—the typical data output per second from recent ONT sequencers. Sequencing latency represents the number of these chunks required to reach a mapping decision for each read, as reported by each tool in its PAF output; lower is better. \uncalled reports the number of bases instead of the number of chunks, which we convert to the number of chunks by dividing by 450. Where available, we report the sequencing latency as the number of bases in \suptabref{full_results}.
    \item \textbf{Accuracy} (F-1 score) based on the annotations from \uncalled's \texttt{pafstats} tool, with \minimap's mapping locations as the ground truth; higher is better. The F-1 score is a common way of combining precision and recall into a single metric, which we report individually in \suptabref{full_results}.
\end{itemize}

\subsubsection{Datasets} \label{sec:datasets}
We use the same datasets as \rawhash's evaluation, labeled as d1 through d5, which include a range of organisms from viruses to humans. Specifically, d1 is SARS-CoV-2, d2 is \textit{E. coli}, d3 is Yeast, d4 is Green Algae, and d5 is Human. We updated the Green Algae dataset (d4) to use all of the available 1.3 Gbp of ERR3237140, whereas \rawhash used only 0.6 Gbp of this dataset. \tabref{datasets} provides detailed information on each dataset.

\begin{table*}[htb]
\centering
    \caption{Dataset Details.}
    \label{table:datasets}
        \begin{tabular}{@{}cllrrllr@{}}\toprule
& \textbf{Organism} & \textbf{Flow Cell} & \textbf{Reads} & \textbf{Bases}              & \textbf{SRA}       & \textbf{Reference}      & \textbf{Genome}\\
&   & \textbf{Version}       & \textbf{(\#)}  & \textbf(\#)               & \textbf{Accession} & \textbf{Genome}         & \textbf{Size}  \\\midrule
\multicolumn{8}{c}{Read Mapping} \\\midrule
d1 & \emph{SARS-CoV-2} & R9.4  & 1,382,016      & 594M                   & CADDE Centre & GCF\_009858895.2   & 29,903 \\\midrule
d2 & \emph{E. coli}    & R9.4 & 353,317        & 2,364M                 & ERR9127551         & GCA\_000007445.1   & 5M \\\midrule
d3 & \emph{Yeast}      & R9.4 & 49,992         & 380M                   & SRR8648503         & GCA\_000146045.2   & 12M\\\midrule
d4 & \emph{Green Algae} & R9.4 & 63,215         & 1,335M                   & ERR3237140         & GCF\_000002595.2   & 111M\\\midrule
d5 & \emph{Human HG001} & R9.4 & 269,507        & 1,584M                 & FAB42260 Nanopore WGS & T2T-CHM13 (v2)     & 3,117M\\\bottomrule
\multicolumn{8}{c}{Relative Abundance Estimation} \\\midrule
\multicolumn{3}{c}{D1-D5} & 2,118,047    & 6,257M                  & d1-d5              & d1-d5              & 3,246M\\\bottomrule
\multicolumn{8}{c}{Contamination Analysis} \\\midrule
\multicolumn{3}{c}{D1 and D5} & 1,651,523    & 2,178M                 & d1 and d5             & d1                 & 29,903\\\bottomrule
\multicolumn{8}{l}{Dataset numbers (\emph{e.g.,} d1-d5) show the combined datasets.}\\
\multicolumn{8}{l}{Datasets are from R9.4. Base counts in millions (M).}\\
\end{tabular}

\end{table*}

\subsubsection{System Specifications}
With two exceptions, we run all CPU evaluations 64 threads on a dual-socket Intel Xeon Gold 6226R (2\x~16 physical cores, 2\x~32 logical cores) at 2.9GHz with 256GB of DDR4 RAM. \sigmap runs out of memory on the Intel system for the Human and Relative Abundance datasets; hence, we run \sigmap with these datasets using 64 threads on a dual-socket AMD EPYC 7742 (2\x~64 physical cores, 2\x~128 logical cores) at 2.25GHz with 1024GB of DDR4 RAM.

\subsubsection{Parameter Settings}
\rev{We empirically select all parameters used in \toolname to optimize performance and accuracy. Specifically, for the \defn{match bonus} $B_{Match}$ and the \defn{score threshold} $Score_{Threshold}$ (discussed in \secref{mappingdecisions}), we set the default values to $B_{Match}=0.4$ and $Score_{Threshold}=20$. These values were determined through a two-dimensional parameter sweep on the d2 \emph{E. coli} dataset to achieve the best overall accuracy. To provide insights about the effect of the band width on the accuracy, we provide our parameter sweep evaluation in \secref{resbwsweep}.}

\subsection{Read Mapping} \label{sec:resreadmapping}
We run a read mapping experiment for all tools and datasets. \tabref{numeric_results} shows the raw numeric results for five key metrics: memory footprint, throughput, analysis latency, sequencing latency, and accuracy, with the best tool in each dataset and metric highlighted in bold. \figref{spidertradeoffs} shows the same results as a spider chart for each dataset, with each metric in a dataset normalized from worst (center, \emph{e.g.,} lowest throughput) to best (outside, \emph{e.g.,} highest throughput). Thus, a tool with a large area in the spider chart indicates a good overall tradeoff in all metrics, while a small area indicates a bad overall tradeoff between the metrics. We omit the \rawhashtwo and \toolname-RH2 results from \figref{spidertradeoffs} for readability.

We make seven key observations:
First, \toolname is consistently highly accurate for all datasets, and it is the only tool that is highly accurate for large reference databases (\emph{e.g.,} Human, Relative Abundance).
Second, \toolname has a low analysis latency (\emph{i.e.,} time taken for computation) and sequencing latency (\emph{i.e.,} time taken for sequencing).
Third, \toolname has a large area in \figref{spidertradeoffs} for all datasets, \emph{i.e.,} \toolname generalizes well to all evaluated datasets.
Fourth, \toolname's memory footprint depends mainly on the reference database size, requiring up to 83GB for the largest reference database (relative abundance), a size typically available in moderately sized servers. For the evaluated datasets with relatively small reference databases (d1-d4, Contamination), the memory footprint is at most 12.2GB, a size typically available in modern desktop computers and some laptops.
Fifth, \toolname's mean throughput per thread is at least that of recent ONT nanopores (450 bp/s), \emph{i.e.,} a single thread can analyze the signals generated by a nanopore in real-time.
Sixth, for some datasets, \toolname has higher throughput than \rawhash, despite having more computational steps. The reason is that \toolname's higher accuracy enables higher confidence and hence quicker decisions to stop sequencing individual reads than in \rawhash, leading to overall less data being analyzed.
Seventh, \toolname generalizes well to other seeding and filtering mechanisms, such as \rawhashtwo, improving accuracy while maintaining high throughput without fine-tuning. \toolname achieves even higher accuracy than \toolname-RH2, demonstrating that fine-tuning the specific combination of seeding, filtering, and alignment steps can further improve accuracy.

We conclude that \toolname is the only tool to generalize well across a wide range of reference database sizes and read set compositions. It is consistently highly accurate, has low latency, meets the real-time analysis throughput requirement of 450bp/s, and fits in the memory of typical laptop, desktop, or moderately sized server systems, depending on the reference database size.

\begin{table}
    \caption{Numeric Results.}
    \label{table:numeric_results}
    \setlength\tabcolsep{3 pt}
    \resizebox{\columnwidth}{!}{%
        \begin{tabular}{l r r r r r}
\hline
\multicolumn{1}{l}{} & \multicolumn{1}{c}{\textbf{Memory}} & \multicolumn{1}{c}{\textbf{Throughput}} & \multicolumn{1}{c}{\textbf{Analysis}} & \multicolumn{1}{c}{\textbf{Sequencing}} & \multicolumn{1}{c}{\textbf{Accuracy}}\\
\multicolumn{1}{l}{\textbf{d1 SARS-CoV-2}} & \multicolumn{1}{c}{\textbf{Footprint (GB)}} & \multicolumn{1}{c}{\textbf{(bp/s)}} & \multicolumn{1}{c}{\textbf{Latency (ms)}} & \multicolumn{1}{c}{\textbf{Latency (Chunks)}} & \multicolumn{1}{c}{\textbf{(F-1)}}\\
\hline
UNCALLED & \textbf{0.28} & 6,575 & 29.24 & \textbf{0.410} & \textbf{0.972}\\
Sigmap & 28.25 & 350,565 & 1.11 & 1.005 & 0.711\\
RawHash & 4.21 & 502,043 & 0.94 & 1.238 & 0.925\\
RawAlign & 4.52 & 438,090 & 1.07 & 1.126 & 0.939\\
RawHash2 & 4.32 & \textbf{670,152} & \textbf{0.71} & 1.219 & 0.935\\
RawAlign-RH2 & 4.31 & 641,164 & 0.74 & 1.266 & 0.917\\
\hline
\multicolumn{2}{l}{\textbf{d2 \emph{E. Coli}}}\\
\hline
UNCALLED & \textbf{0.80} & 5,174 & 115.79 & \textbf{1.290} & \textbf{0.973}\\
Sigmap & 111.17 & 19,216 & 34.44 & 2.111 & 0.967\\
RawHash & 4.27 & 49,560 & 19.75 & 3.200 & 0.928\\
RawAlign & 4.31 & 54,263 & 13.11 & 1.995 & 0.968\\
RawHash2 & 4.46 & 128,692 & 7.51 & 3.270 & 0.934\\
RawAlign-RH2 & 4.45 & \textbf{132,710} & \textbf{6.35} & 2.773 & 0.943\\
\hline
\multicolumn{2}{l}{\textbf{d3 Yeast}}\\
\hline
UNCALLED & \textbf{0.58} & 5,152 & 159.30 & \textbf{2.773} & 0.941\\
Sigmap & 14.71 & 15,217 & 67.60 & 4.139 & 0.947\\
RawHash & 4.53 & 17,997 & 77.59 & 5.826 & 0.906\\
RawAlign & 4.53 & 17,855 & 48.39 & 3.071 & \textbf{0.963}\\
RawHash2 & 4.83 & 51,912 & 24.95 & 4.263 & 0.921\\
RawAlign-RH2 & 4.89 & \textbf{55,515} & \textbf{18.63} & 3.321 & 0.944\\
\hline
\multicolumn{2}{l}{\textbf{d4 Green Algae}}\\
\hline
UNCALLED & \textbf{1.26} & \textbf{8,174} & 440.81 & 11.790 & 0.840\\
Sigmap & 53.71 & 2,251 & 608.90 & 5.804 & \textbf{0.938}\\
RawHash & 14.06 & 5,430 & 700.30 & 10.646 & 0.824\\
RawAlign & 12.20 & 5,871 & 276.09 & 4.514 & 0.932\\
RawHash2 & 8.77 & 6,176 & 269.97 & 5.025 & 0.904\\
RawAlign-RH2 & 9.45 & 6,897 & \textbf{210.83} & \textbf{4.064} & 0.910\\
\hline
\multicolumn{2}{l}{\textbf{d5 Human}}\\
\hline
UNCALLED & \textbf{13.17} & \textbf{5,613} & \textbf{1,077.54} & 12.959 & 0.320\\
Sigmap & 313.40 & 195 & 16,296.43 & 10.401 & 0.327\\
RawHash & 56.94 & 1,299 & 6,318.98 & 10.695 & 0.557\\
RawAlign & 80.35 & 886 & 3,721.86 & \textbf{6.321} & \textbf{0.702}\\
RawHash2 & 96.39 & 540 & 6,357.51 & 7.794 & 0.585\\
RawAlign-RH2 & 119.29 & 391 & 6,133.61 & 7.126 & 0.642\\
\hline
\multicolumn{2}{l}{\textbf{Contamination}}\\
\hline
UNCALLED & \textbf{1.06} & 6,608 & 199.28 & 3.557 & \textbf{0.964}\\
Sigmap & 111.65 & 405,956 & 1.21 & 2.062 & 0.650\\
RawHash & 4.28 & 524,043 & 1.14 & 2.409 & 0.872\\
RawAlign & 4.50 & 455,376 & 2.00 & 3.227 & 0.938\\
RawHash2 & 4.26 & 644,209 & \textbf{0.87} & \textbf{1.659} & 0.885\\
RawAlign-RH2 & 4.24 & \textbf{655,537} & 1.15 & 1.796 & 0.917\\
\hline
\multicolumn{2}{l}{\textbf{Relative Abundance}}\\
\hline
UNCALLED & \textbf{10.87} & \textbf{6,722} & \textbf{309.08} & 4.921 & 0.218\\
Sigmap & 506.34 & 182 & 5,670.36 & 3.338 & 0.406\\
RawHash & 60.76 & 597 & 2,264.01 & 3.816 & 0.459\\
RawAlign & 83.76 & 480 & 1,652.16 & \textbf{2.336} & \textbf{0.754}\\
RawHash2 & 97.56 & 562 & 2,122.52 & 3.404 & 0.363\\
RawAlign-RH2 & 119.02 & 576 & 1,885.22 & 3.131 & 0.418\\
\hline
\end{tabular}

    }
\end{table}

\subsection{Relative Abundance} \label{sec:resrelativeabundance}
The relative abundance read set is artificially composed of the read sets from d1-d5; hence, the ground truth is known. We map the read set against the concatenation of the reference genomes d1-d5 and calculate relative abundance estimates based on the number of reads per organism reported by each tool. We consider \minimap with basecalled reads as input as an additional baseline. We use the Euclidean distance to the ground truth relative abundance as the main accuracy metric. \tabref{read_ratio_relative_abundances} shows the results. We report the relative abundance results based on the number of bases instead of the number of reads in \supsecref{bases_ratio_relative_abundance}.

We make three key observations. First, \toolname is the most accurate among the raw signal analysis tools with a Euclidean distance of 0.12 to the ground truth and close to \minimap's accuracy with a Euclidean distance of 0.05 to the ground truth. Second, \toolname is the most accurate among the raw signal analysis tools for all organisms and is within 0.07 of the Euclidean distance of \minimap. Note that \minimap has the advantage of having full-length basecalled reads as input, while \toolname uses only a short prefix of each read for mapping.
Third, we observe that \toolname is much more accurate than \toolname-RH2. The reason is that \toolname-RH2 drops some correct candidates during the earlier aggressive \defn{frequency filtering} of seeds~\cite{firtina_rawhash2_2024}, where high-frequency seeds that occur too often in the reference are discarded to reduce computational load. However, this can eliminate valid matches, which cannot be recovered by the later alignment step, potentially leading to biased relative abundance estimates. In contrast, \toolname does not apply frequency filtering, resulting in high accuracy at the cost of some performance (see \tabref{numeric_results}).

We draw two key takeaways. First, \toolname can be directly used to estimate relative abundances as an end-to-end use case with high accuracy and without the need for basecalling. Second, early aggressive filters can limit the benefits of alignment. Fine-tuning filters leads to better accuracy/performance tradeoffs.

\begin{table}%
    \caption{Read Ratio Relative Abundances.}
    \label{table:read_ratio_relative_abundances}
    \resizebox{\columnwidth}{!}{%
        \begin{tabular}{l r r r r r r}
\hline
\textbf{Tool} & \textbf{SARS-CoV-2} & \emph{\textbf{E. coli}} & \textbf{Yeast} & \textbf{Green Algae} & \textbf{Human} & \textbf{Distance}\\
\hline
Ground Truth & 0.652 & 0.167 & 0.024 & 0.030 & 0.127 & -\\
minimap2 & 0.613 & 0.163 & 0.025 & 0.053 & 0.147 & \textbf{0.050}\\
UNCALLED & 0.072 & 0.466 & 0.001 & 0.150 & 0.312 & 0.689\\
Sigmap & 0.201 & 0.446 & 0.002 & 0.123 & 0.229 & 0.549\\
RawHash & 0.309 & 0.440 & 0.000 & 0.073 & 0.178 & 0.445\\
RawAlign & 0.565 & 0.248 & 0.002 & 0.050 & 0.136 & 0.123\\
RawHash2 & 0.130 & 0.537 & 0.000 & 0.123 & 0.210 & 0.653\\
RawAlign-RH2 & 0.208 & 0.491 & 0.000 & 0.099 & 0.203 & 0.560\\
\hline
\end{tabular}

    }
\end{table}

\subsection{Band Width Parameter Sweep} \label{sec:resbwsweep}
\begin{figure}[htb]
  \centering
    \includegraphics[width=\linewidth]{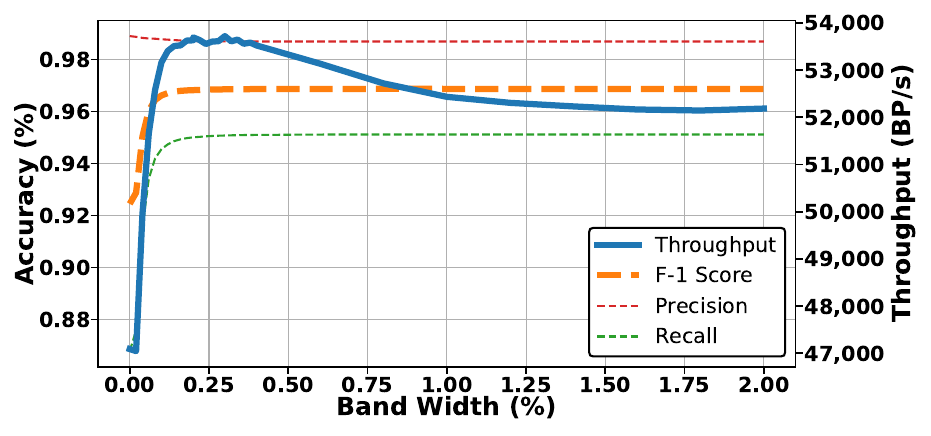}
    \hfill
  \caption{Band Width Parameter Sweep on d2 \textit{E. coli}.}
  \label{fig:bandwidthsweep}
\end{figure}

To minimize the computational overhead of alignment, we use banded dynamic time warping (see \secref{banded_alignment}). If the width of the band is chosen too narrow, accuracy will be degraded. If it is too wide, performance will suffer. We parameter sweep the band width as a fraction of the length of the read fragment to be aligned on the d2 \emph{E. Coli} dataset, and plot accuracy and throughput. \figref{bandwidthsweep} shows the results.

We observe that a band width of 20\% of the length of the read fragment is sufficient to achieve the best accuracy (in terms of F-1 score), while simultaneously maximizing throughput. The reason that throughput is also maximized is that the pathological behavior of a too-narrow band width causes low confidence mapping candidates, meaning reads have to be analyzed for longer to reach a decision.

Based on this parameter sweep, we choose 20\% as the default band width for \toolname.

\subsection{DTW Performance Optimizations}\label{sec:optimizationimpact}
\begin{figure}[htb]
  \centering
    \includegraphics[width=0.9\linewidth]{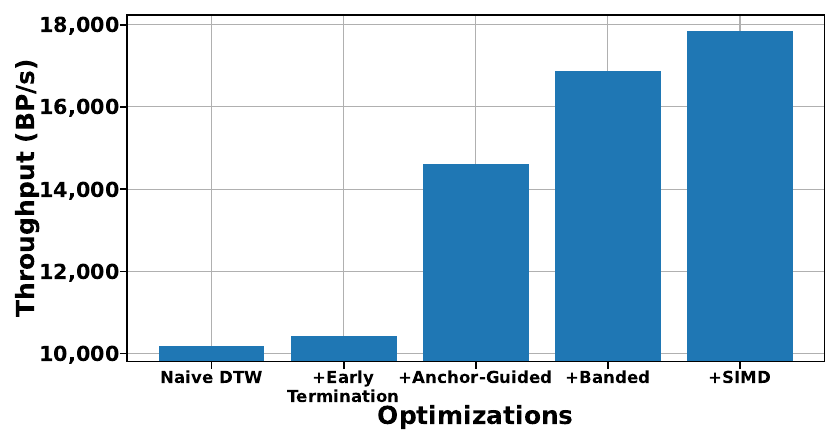}
    \hfill
  \caption{Throughput with each Additional DTW Performance Optimization on d3 Yeast.}
  \label{fig:optimizations}
\end{figure}

To understand the impact of each performance optimization proposed in \secref{perfopt}, we run a read mapping experiment on the d3 Yeast dataset. The experiment starts with a \defn{naive DTW implementation}, an unoptimized version of DTW that computes the full dynamic programming matrix without any performance enhancements. In each subsequent run, we iteratively enable a new optimization on top of the previous ones. \figref{optimizations} shows the results.

We make three key observations. First, we observe that anchor-guided alignments, banded alignments, and SIMD each significantly increase throughput.   Second, we observe that the early termination strategy yields only a small benefit on the d3 Yeast dataset. This is mainly due to the relatively small reference genome size, which results in a small number of candidate regions per read. In contrast, the early termination strategy is most powerful when it can quickly eliminate many poor candidates of a read after a first good candidate is found. Thus, the early termination strategy is strongest for large reference databases. Third, we observe that in the case of the d3 Yeast dataset, even a naive DTW implementation has a reasonably high throughput. This is because although the naive DTW subroutine is inefficient, it is called infrequently due to the small number of candidate regions per read. This is not the case for large reference databases.

We conclude that (1)~the performance optimizations presented in \secref{perfopt} are effective, and (2)~their relative importance depends on the size of the reference database.

\section{Related Work} \label{sec:rel_work}
Several studies have explored real-time genome analysis of raw nanopore signals by leveraging adaptive sampling techniques~\cite{\citesignalanalysis}. Methods such as SquiggleNet~\cite{bao_squigglenet_2021}, DeepSelectNet~\cite{senanayake2023deepselectnet}, and RawMap~\cite{sadasivan_rapid_2023} employ machine learning techniques to classify raw nanopore signals to specific species without performing read mapping. Sigmoni~\cite{shivakumar_sigmoni_2024} adopts a strategy similar to that of SPUMONI~\cite{ahmed_pan-genomic_2021} and SPUMONI 2~\cite{ahmed_spumoni_2023}, facilitating classification without basecalling.

UNCALLED~\cite{kovaka_targeted_2021}, Sigmap~\cite{zhang_real-time_2021}, RawHash~\cite{firtina2023rawhash}, and RawHash2~\cite{firtina_rawhash2_2024} are particularly relevant to our work, as they map raw nanopore signals to a reference genome without relying on computationally intensive basecalling steps.

UNCALLED~\cite{kovaka_targeted_2021} maps raw signals to a reference genome by computing the probabilities of k-mers that each raw signal segment (\emph{i.e.,} event) may represent, using a k-mer model that provides expected event values for all possible k-mers. It then identifies sequences of matching k-mers between the most probable k-mers of events and the reference genome using an FM-index~\cite{ferragina_opportunistic_2000}. However, as the genome size increases, this probabilistic model faces challenges in accurately identifying matching regions due to the vast number of potential matches~\cite{kovaka_targeted_2021}. While UNCALLED maintains high accuracy for small genomes (\emph{e.g.,} \emph{E. coli} and \emph{Yeast}), its accuracy significantly declines when applied to larger genomes, such as the human genome.

Sigmap~\cite{zhang_real-time_2021} addresses mapping to larger genomes by computing the Euclidean distance between raw signals and a reference genome, after converting the reference genome into its expected signal representation using k-mer models. This distance calculation allows for identifying segments of raw signals that closely match particular regions in the reference genome. However, calculating the Euclidean distance between high-dimensional vectors is computationally expensive and suffers from the \emph{curse of dimensionality}~\cite{bellman_dynamic_1966}, limiting its scalability and efficiency for large genomes.

RawHash~\cite{firtina2023rawhash} maps raw nanopore signals to a reference genome using a quick hash-based search followed by chaining. To do so, RawHash quantized (\emph{i.e.,} bucketing) noisy raw signals such that the signals from the same DNA content are assigned to the same hash value even though their corresponding raw signals may be slightly different than each other. This coarse-grained hash-based sequence comparison is fast and scales well to large reference genomes but has relatively low accuracy. RawHash2~\cite{firtina_rawhash2_2024} provides a more sensitive quantization mechanism and mapping strategies. In doing so, RawHash2 provides improvements in terms of speedup and accuracy compared to RawHash. However, RawHash and RawHash2 still lack fine-grained sequence alignments in the output, limiting their applicability to downstream analysis tasks (\emph{e.g.,} variant calling).

Although earlier works~\cite{dunn2021squigglefilter, shih_efficient_2022, sadasivan2023accelerating, gamaarachchi2020dtwgpu} have explored performing alignment on raw nanopore signals using DTW~\cite{\dtwSeminal}, these works lack an accurate and fast seeding mechanism. Without such a seeding mechanism, costly alignment operations must be performed for every position in the reference genome. This approach is impractical as it cannot scale to larger genomes, especially when performing real-time analysis. \toolname is the first work to integrate a fast and accurate hash-based seeding mechanism~\cite{firtina2023rawhash, firtina_rawhash2_2024} before performing the alignment. By doing so, \toolname makes it practical to perform alignment on larger genomes.

\section{Discussion} \label{sec:discussion}
\toolname's Seed-Filter-Align approach is inspired by conventional basecalled read mappers. We have compared \toolname to the highly engineered state-of-the-art basecalled read mapper \minimap in the relative abundance use case and observe that \toolname is approaching the accuracy of \minimap. In the future, developing downstream analysis tools that operate directly on raw nanopore signals may enable more insightful and direct comparisons between analyses conducted on basecalled sequences and those performed on raw signal data. Such tools could reveal nuances and variations in the data often lost during the basecalling process, potentially leading to an improved understanding of genomic variations and more accurate genomic analyses. \rev{To achieve this, we identify two key limitations that can be tackled in future work to improve the accuracy and performance of raw signal analysis. 

First, \toolname cannot perform alignment between a pair of raw signals where none of the pairs are from a reference genome. This is mainly due to the seeding step \toolname builds on and \toolname's requirement to perform segmentation before performing the alignment (\emph{i.e.,} the alignment is performed between events). Performing a pairwise alignment between raw nanopore signals can enable identifying overlaps between signals and subsequently constructing \emph{de novo} assemblies from these overlaps as miniasm does~\cite{li_minimap_2016}.

Second, the specific alignment operations (\emph{i.e.,} edit operations such as insertions, deletions, and substitutions) between a raw nanopore signal and a reference genome can be identified by performing \emph{backtracking} after constructing the DTW dynamic programming table. Backtracking involves tracing the optimal path through the dynamic programming matrix from the endpoint back to the start, allowing us to reconstruct the sequence of edits that constitute the optimal alignment. Generating these edit operations in a mapping file can be useful to enable subsequent steps in genome analysis, such as variant calling. However, backtracking significantly increases computational overhead because it requires additional memory and processing time to trace the optimal alignment path through the DTW matrix. To enable \toolname to scale to much larger genomes while generating edit operations in real-time, further algorithmic optimizations are needed. These may include more efficient data structures for storing the dynamic programming matrix, heuristic methods to approximate the alignment path, or parallelization techniques to accelerate backtracking.}

We find that Seed-Filter-Align is already an effective paradigm for raw signal mapping as implemented in \toolname, and in the future, may directly compete with basecalled read mappers once it is engineered to a similar extent as existing basecalled read mappers.

\section{Conclusion} \label{sec:conclusion}
We present \toolname, a tool that improves the state-of-the-art in raw nanopore signal mapping by integrating fast and scalable hash-based seeding with highly accurate alignment using dynamic time warping (DTW). We address key challenges in combining DTW with hash-based mapping, including computational cost and integration into the mapping decision process, by proposing optimizations such as chaining-based filtering, early termination, anchor-guided alignment, and SIMD vectorization.

To our knowledge, \toolname is the first work to combine raw signal seeding and alignment via dynamic time warping. We have extensively compared \toolname to \uncalled~\cite{kovaka_targeted_2021}, \sigmap~\cite{zhang_real-time_2021}, \rawhash~\cite{firtina2023rawhash}, and \rawhashtwo~\cite{firtina_rawhash2_2024}. In our comparison, we have demonstrated the benefits of \toolname's Seed-Filter-Align approach over prior methods for raw signal mapping.

Our comprehensive evaluations demonstrate that \toolname consistently achieves high accuracy and low latency across a wide range of reference genome sizes, including large genomes like the human genome. \toolname meets the real-time analysis throughput requirements and generalizes well to different datasets. We conclude that \toolname is the first tool to enable accurate, fast, and scalable mapping of raw nanopore signals to large genomes without basecalling, making it practical for real-time analysis and downstream tasks such as variant calling.

\section*{Acknowledgments}
SAFARI Research Group acknowledges the generous gift funding provided by our industrial partners (especially by Google, Huawei, Intel, Microsoft, VMware), which has been instrumental in enabling the 15+ year-long research SAFARI Research Group has been conducting on accelerating genome analysis. This work is also partially supported by the Semiconductor Research Corporation (SRC), the European Union’s Horizon programme for research and innovation [101047160 - BioPIM] and the Swiss National Science Foundation (SNSF) [200021\_213084].

\bibliographystyle{IEEEtran}
\setstretch{0.75}
{\small \bibliography{main}}

\onecolumn
\setcounter{secnumdepth}{3}
\clearpage
\begin{center}
\textbf{\LARGE Supplementary Material for\\ \fulltitle}
\end{center}
\setcounter{section}{0}
\setcounter{equation}{0}
\setcounter{figure}{0}
\setcounter{table}{0}
\setcounter{page}{1}
\makeatletter
\renewcommand{\theequation}{S\arabic{equation}}
\renewcommand{\thetable}{S\arabic{table}}
\renewcommand{\thefigure}{S\arabic{figure}}
\renewcommand{\thesection}{\Alph{section}}
\renewcommand{\thesubsection}{\thesection.\arabic{subsection}}
\renewcommand{\thesubsubsection}{\thesubsection.\arabic{subsubsection}}

\section{Background}
\subsection{Offline Reference Database Pre-Processing}\label{supsec:ref_preprocessing}
A \defn{reference database} is a set of a priori known biological sequences that reads should be compared to during mapping. For example, the reference database could consist of a chromosome, a set of genes, a genome, or multiple
genomes. Reference databases are often available and accepted by read mappers in FASTA file format, \emph{i.e.,} as a set of basecalled nucleotide sequences. \toolname pre-processes the reference database in two steps: First, basecalled nucleotide reference sequences are converted into a form that is easier to compare with a raw signal read~(\supsecref{expected_event_generation}). Second, the reference sequences are \defn{indexed} into a hash table that allows querying for small exact matches during mapping~(\supsecref{indexing}).

\subsubsection{Expected Event Generation}\label{supsec:expected_event_generation}
To ease comparison with a raw signal read, \toolname converts the basecalled reference database into sequences of raw signals that would be observed in a perfect noise-free sequencer, called \defn{expected event sequences}. \toolname obtains expected event sequences in an offline pre-processing step based on a model of the nanopore provided by the manufacturer, like in prior works (\emph{e.g.,}~\citesupp{supp_firtina2023rawhash, supp_zhang_real-time_2021}).

\subsubsection{Indexing}\label{supsec:indexing}
Mapping with the Seed-Filter-Align paradigm requires an \defn{index}, a data structure that can be efficiently queried for locations of small exact matches between the read and the reference. To this end, \toolname cuts the expected event sequences into short snippets and quantizes them into \defn{seeds}. The seeds are inserted into a hash table, where each seed is a key, and the value is a list of locations where that seed occurs in the reference database. \toolname leverages the pre-processing steps from \rawhash~\citesupp{supp_firtina2023rawhash}.

\subsection{Online Read Pre-Processing}\label{supsec:read_preprocessing}
Current nanopore sequencers take multiple measurements as each nucleotide base passes through the nanopore, leading to many largely redundant measurements. To simplify the further computation, \toolname converts them into \defn{event sequences}, where redundant measurements are averaged into a single value. This conversion is called \defn{segmentation}. It is not trivial since the translocation rate of the sequence molecule through the nanopore is variable, \emph{i.e.,} it is not obvious which groups of signals should be averaged together.
\toolname leverages the segmentation logic proposed in Scrappie~\citesupp{supp_scrappie} and used by \uncalled~\citesupp{supp_kovaka_targeted_2021}, \sigmap~\citesupp{supp_zhang_real-time_2021}, and \rawhash~\citesupp{supp_firtina2023rawhash}.

\clearpage
\section{Full Results}
\suptabref{full_results} reports the full results of the read mapping evaluation in \secref{resreadmapping} with the best tool in each dataset and metric highlighted in bold. The metrics are defined as follows:

\begin{itemize}
    \item \textbf{Indexing memory footprint} (GB) during indexing as reported by Ubuntu's \texttt{time} package; lower is better.
    \item \textbf{Mapping memory footprint} (GB) during mapping as reported by Ubuntu's \texttt{time} package; lower is better.
    \item \textbf{Mean throughput} (bp/s) per thread of mapping as reported by each tool in its PAF output. The throughput should match at least the throughput of a nanopore (\emph{e.g.,} 450 bp/s in recent ONT sequencers); higher is better.
    \item \textbf{Median throughput} (bp/s) per thread of mapping as reported by each tool in its PAF output. The throughput should match at least the throughput of a nanopore (\emph{e.g.,} 450 bp/s in recent ONT sequencers); higher is better.
    \item Mean \textbf{analysis latency} (ms), \emph{i.e.,} the time spent computing for mapping reads as reported by each tool in its PAF output; lower is better. We also report the median analysis latency in \suptabref{full_results}.
    \item Mean \textbf{sequencing latency} (bases), \emph{i.e.,} the number of bases needed to reach a mapping decision for each read, as reported by each tool in its PAF output; lower is better.
    \item Mean \textbf{sequencing latency} (chunks), \emph{i.e.,} the number of 450 basepair-sized chunks needed to reach a mapping decision for each read, as reported by each tool in its PAF output; lower is better. Uncalled reports a number of bases instead of a number of chunks, which we convert to a number of chunks by dividing by 450.
    \item \textbf{Precision} based on the annotations from Uncalled's \texttt{pafstats} tool, with minimap2's mapping locations as the ground truth; higher is better.
    \item \textbf{Recall} based on the annotations from Uncalled's \texttt{pafstats} tool, with minimap2's mapping locations as the ground truth; higher is better.
    \item \textbf{F-1} score based on the annotations from Uncalled's \texttt{pafstats} tool, with minimap2's mapping locations as the ground truth; higher is better. The F-1 score is a combination of precision and recall.
\end{itemize}

\begin{sidewaystable*}
    \setlength\tabcolsep{3 pt}

    \caption{Numeric Results.}
    \label{suptable:full_results}
\resizebox{\textwidth}{!}{%
    \begin{tabular}{l r r r r r r r r r r}
\hline
\multicolumn{1}{l}{} & \multicolumn{1}{c}{\textbf{Indexing Memory}} & \multicolumn{1}{c}{\textbf{Mapping Memory}} & \multicolumn{1}{c}{\textbf{Mean Throughput}} & \multicolumn{1}{c}{\textbf{Median Throughput}} & \multicolumn{1}{c}{\textbf{Analysis}} & \multicolumn{1}{c}{\textbf{Mean Sequencing}} & \multicolumn{1}{c}{\textbf{Mean Sequencing}} & \multicolumn{1}{c}{} & \multicolumn{1}{c}{} & \multicolumn{1}{c}{}\\
\multicolumn{1}{l}{\textbf{d1 SARS-CoV-2}} & \multicolumn{1}{c}{\textbf{Footprint (GB)}} & \multicolumn{1}{c}{\textbf{Footprint (GB)}} & \multicolumn{1}{c}{\textbf{(bp/s)}} & \multicolumn{1}{c}{\textbf{(bp/s)}} & \multicolumn{1}{c}{\textbf{Latency (ms)}} & \multicolumn{1}{c}{\textbf{Latency (Bases)}} & \multicolumn{1}{c}{\textbf{Latency (Chunks)}} & \multicolumn{1}{c}{\textbf{Precision}} & \multicolumn{1}{c}{\textbf{Recall}} & \multicolumn{1}{c}{\textbf{F-1}}\\
\hline
UNCALLED & 0.07 & \textbf{0.28} & 6,575 & 6,433 & 29.24 & \textbf{185} & \textbf{0.410} & 0.955 & \textbf{0.991} & \textbf{0.972}\\
Sigmap & \textbf{0.01} & 28.25 & 350,565 & 355,001 & 1.11 & - & 1.005 & 0.993 & 0.554 & 0.711\\
RawHash & \textbf{0.01} & 4.21 & 502,043 & 454,272 & 0.94 & 514 & 1.238 & 0.983 & 0.874 & 0.925\\
RawAlign & \textbf{0.01} & 4.52 & 438,090 & 425,341 & 1.07 & - & 1.126 & 1.000 & 0.885 & 0.939\\
RawHash2 & \textbf{0.01} & 4.32 & \textbf{670,152} & \textbf{662,431} & \textbf{0.71} & 498 & 1.219 & 0.977 & 0.896 & 0.935\\
RawAlign-RH2 & \textbf{0.01} & 4.31 & 641,164 & 639,757 & 0.74 & 522 & 1.266 & \textbf{1.000} & 0.846 & 0.917\\
\hline
\multicolumn{2}{l}{\textbf{d2 \emph{E. Coli}}}\\
\hline
UNCALLED & \textbf{0.12} & \textbf{0.80} & 5,174 & 5,115 & 115.79 & \textbf{580} & \textbf{1.290} & 0.982 & \textbf{0.965} & \textbf{0.973}\\
Sigmap & 0.40 & 111.17 & 19,216 & 18,063 & 34.44 & - & 2.111 & 0.984 & 0.950 & 0.967\\
RawHash & 0.35 & 4.27 & 49,560 & 44,379 & 19.75 & 1,376 & 3.200 & 0.956 & 0.901 & 0.928\\
RawAlign & 0.39 & 4.31 & 54,263 & 48,778 & 13.11 & - & 1.995 & 0.987 & 0.950 & 0.968\\
RawHash2 & 0.35 & 4.46 & 128,692 & 124,544 & 7.51 & 1,380 & 3.270 & 0.986 & 0.887 & 0.934\\
RawAlign-RH2 & 0.39 & 4.45 & \textbf{132,710} & \textbf{129,523} & \textbf{6.35} & 1,163 & 2.773 & \textbf{0.989} & 0.901 & 0.943\\
\hline
\multicolumn{2}{l}{\textbf{d3 Yeast}}\\
\hline
UNCALLED & \textbf{0.30} & \textbf{0.58} & 5,152 & 4,671 & 159.30 & \textbf{1,248} & \textbf{2.773} & 0.944 & 0.937 & 0.941\\
Sigmap & 1.04 & 14.71 & 15,217 & 14,524 & 67.60 & - & 4.139 & \textbf{0.986} & 0.911 & 0.947\\
RawHash & 0.76 & 4.53 & 17,997 & 16,853 & 77.59 & 2,566 & 5.826 & 0.985 & 0.839 & 0.906\\
RawAlign & 0.86 & 4.53 & 17,855 & 16,369 & 48.39 & - & 3.071 & 0.962 & \textbf{0.964} & \textbf{0.963}\\
RawHash2 & 0.76 & 4.83 & 51,912 & 50,323 & 24.95 & 1,825 & 4.263 & 0.963 & 0.882 & 0.921\\
RawAlign-RH2 & 0.86 & 4.89 & \textbf{55,515} & \textbf{53,853} & \textbf{18.63} & 1,408 & 3.321 & 0.958 & 0.930 & 0.944\\
\hline
\multicolumn{2}{l}{\textbf{d4 Green Algae}}\\
\hline
UNCALLED & 11.96 & \textbf{1.26} & \textbf{8,174} & \textbf{8,132} & 440.81 & 5,306 & 11.790 & 0.888 & 0.798 & 0.840\\
Sigmap & 8.63 & 53.71 & 2,251 & 2,150 & 608.90 & - & 5.804 & \textbf{0.974} & 0.905 & \textbf{0.938}\\
RawHash & 5.33 & 14.06 & 5,430 & 4,850 & 700.30 & 4,723 & 10.646 & 0.966 & 0.719 & 0.824\\
RawAlign & 6.16 & 12.20 & 5,871 & 5,238 & 276.09 & - & 4.514 & 0.913 & \textbf{0.953} & 0.932\\
RawHash2 & \textbf{5.26} & 8.77 & 6,176 & 5,825 & 269.97 & 2,157 & 5.025 & 0.936 & 0.874 & 0.904\\
RawAlign-RH2 & 6.09 & 9.45 & 6,897 & 6,623 & \textbf{210.83} & \textbf{1,734} & \textbf{4.064} & 0.925 & 0.896 & 0.910\\
\hline
\multicolumn{2}{l}{\textbf{d5 Human}}\\
\hline
UNCALLED & \textbf{48.43} & \textbf{13.17} & \textbf{5,613} & \textbf{6,064} & \textbf{1,077.54} & 5,832 & 12.959 & 0.487 & 0.238 & 0.320\\
Sigmap & 227.77 & 313.40 & 195 & 129 & 16,296.43 & - & 10.401 & 0.429 & 0.264 & 0.327\\
RawHash & 83.09 & 56.94 & 1,299 & 269 & 6,318.98 & 4,773 & 10.695 & \textbf{0.894} & 0.405 & 0.557\\
RawAlign & 107.44 & 80.35 & 886 & 291 & 3,721.86 & - & \textbf{6.321} & 0.696 & \textbf{0.707} & \textbf{0.702}\\
RawHash2 & 83.06 & 96.39 & 540 & 302 & 6,357.51 & 3,441 & 7.794 & 0.877 & 0.439 & 0.585\\
RawAlign-RH2 & 107.43 & 119.29 & 391 & 331 & 6,133.61 & \textbf{3,139} & 7.126 & 0.874 & 0.507 & 0.642\\
\hline
\multicolumn{2}{l}{\textbf{Contamination}}\\
\hline
UNCALLED & 0.07 & \textbf{1.06} & 6,608 & 6,431 & 199.28 & 1,601 & 3.557 & 0.938 & \textbf{0.991} & \textbf{0.964}\\
Sigmap & \textbf{0.01} & 111.65 & 405,956 & 364,466 & 1.21 & - & 2.062 & 0.786 & 0.554 & 0.650\\
RawHash & \textbf{0.01} & 4.28 & 524,043 & 461,281 & 1.14 & 1,039 & 2.409 & 0.870 & 0.874 & 0.872\\
RawAlign & \textbf{0.01} & 4.50 & 455,376 & 433,378 & 2.00 & - & 3.227 & 0.997 & 0.885 & 0.938\\
RawHash2 & \textbf{0.01} & 4.26 & 644,209 & 638,549 & \textbf{0.87} & \textbf{682} & \textbf{1.659} & 0.873 & 0.896 & 0.885\\
RawAlign-RH2 & \textbf{0.01} & 4.24 & \textbf{655,537} & \textbf{645,277} & 1.15 & 768 & 1.796 & \textbf{1.000} & 0.846 & 0.917\\
\hline
\multicolumn{2}{l}{\textbf{Relative Abundance}}\\
\hline
UNCALLED & \textbf{47.79} & \textbf{10.87} & \textbf{6,722} & \textbf{7,233} & \textbf{309.08} & 2,215 & 4.921 & 0.763 & 0.127 & 0.218\\
Sigmap & 238.32 & 506.34 & 182 & 166 & 5,670.36 & - & 3.338 & 0.793 & 0.273 & 0.406\\
RawHash & 153.12 & 60.76 & 597 & 402 & 2,264.01 & 1,685 & 3.816 & 0.947 & 0.303 & 0.459\\
RawAlign & 163.90 & 83.76 & 480 & 339 & 1,652.16 & - & \textbf{2.336} & 0.946 & \textbf{0.627} & \textbf{0.754}\\
RawHash2 & 153.29 & 97.56 & 562 & 450 & 2,122.52 & 1,495 & 3.404 & 0.950 & 0.224 & 0.363\\
RawAlign-RH2 & 163.89 & 119.02 & 576 & 521 & 1,885.22 & \textbf{1,371} & 3.131 & \textbf{0.961} & 0.267 & 0.418\\
\hline
\end{tabular}

}
\end{sidewaystable*}

\clearpage
\section{Bases Ratio Relative Abundances} \label{supsec:bases_ratio_relative_abundance}
We calculate relative abundance estimates based on the number of bases per organism reported by each tool. This metric corresponds to the evaluation in RawHash~\citesupp{supp_firtina2023rawhash}, but we believe the estimation based on the number of reads per organism is more representative in the context on ReadUntil. The read set is artificially composed of the read sets from d1-d5; hence, the ground truth is known. We consider minimap2 with basecalled reads as input as an additional baseline. We use the Euclidean distance to the ground truth relative abundance as the main accuracy metric. \suptabref{bases_ratio_relative_abundances} shows the results. We report the relative abundance results based on the number of reads instead of the number of bases in \secref{resrelativeabundance}.

We observe that \toolname's estimates are entirely inaccurate, even though it achieves a high F-1 score on the Relative Abundance dataset. The reason is that \toolname stops sequencing each read after successfully mapping between 200-400 bases on average for all organisms (a desired property in the context of ReadUntil) thus it weights each read approximately equally. This is why the bases ratio-relative abundances reported by \toolname are close to the read ratio-relative abundances reported in \secref{resrelativeabundance}. In truth, the reads vary significantly in length, meaning \toolname weighs them "wrongly", although it operates as desired in the context of ReadUntil. It is unclear why all raw signal baselines seemingly weigh reads correctly. Since all baselines have significantly worse F-1 scores on the Relative Abundance dataset, there is a possibility that this is down to good fortune. Note that minimap2 is run on the complete reads and not reads "cut short" due to ReadUntil, hence it can weigh the reads correctly.

We conclude that when estimating relative abundances using \toolname, it is more representative to use the number of reads mapped instead of the number of bases mapped.

\begin{table}[htb]
\centering
\caption{Bases Ratio Relative Abundances.}
\label{suptable:bases_ratio_relative_abundances}
\begin{tabular}{l r r r r r r}
\hline
\textbf{Tool} & \textbf{SARS-CoV-2} & \textbf{E. Coli} & \textbf{Yeast} & \textbf{Green Algae} & \textbf{Human} & \textbf{Distance}\\
\hline
Ground Truth & 0.095 & 0.378 & 0.061 & 0.213 & 0.253 & -\\
minimap2 & 0.081 & 0.383 & 0.061 & 0.227 & 0.248 & \textbf{0.021}\\
UNCALLED & 0.002 & 0.561 & 0.000 & 0.228 & 0.208 & 0.219\\
Sigmap & 0.044 & 0.442 & 0.008 & 0.156 & 0.350 & 0.148\\
RawHash & 0.129 & 0.486 & 0.001 & 0.129 & 0.254 & 0.153\\
RawAlign & 0.501 & 0.255 & 0.003 & 0.064 & 0.176 & 0.460\\
RawHash2 & 0.036 & 0.512 & 0.001 & 0.154 & 0.298 & 0.175\\
RawAlign-RH2 & 0.074 & 0.512 & 0.000 & 0.125 & 0.288 & 0.177\\
\hline
\end{tabular}

\end{table}

\section{Reproducing the Results}\label{supsec:reproducing}
All results can easily be reproduced by following the steps in the README at~\url{https://github.com/CMU-SAFARI/RawAlign/blob/ee59c4e5d04d8464e134fe5c0c9e1cd38ccc9cc3/test/README.md}.

In particular, all experimental results are reproduced in one go by the following commands:
{ \fontsize{8}{10}\selectfont
\begin{lstlisting}[language=bash,frame=single]
#!/bin/bash
#number of threads
THR=64

cd read_mapping
cd d1_sars-cov-2_r94 && bash 0_run_all.sh $THR \
  && cd comparison &&  bash 0_run.sh && cd ../..
cd d2_ecoli_r94 && bash 0_run_all.sh $THR \
  && cd comparison &&  bash 0_run.sh && cd ../..
cd d3_yeast_r94 && bash 0_run_all.sh $THR \
  && cd comparison &&  bash 0_run.sh && cd ../..
cd d4_green_algae_r94 && bash 0_run_all.sh $THR \
  && cd comparison &&  bash 0_run.sh && cd ../..
cd d5_human_na12878_r94 && bash 0_run_all.sh $THR \
  && cd comparison &&  bash 0_run.sh && cd ../..
cd ..

cd relative_abundance && \
  bash 0_run_all.sh $THR && cd ..
cd contamination && \
  bash 0_run_all.sh $THR && cd ..
\end{lstlisting}
}

\let\noopsort\undefined
\let\printfirst\undefined
\let\singleletter\undefined
\let\switchargs\undefined

\bibliographystylesupp{IEEEtran}
\setstretch{0.75}
{\small \bibliographysupp{supp}}

\end{document}